# Dynamics of nano-scale assemblies of amphiphilic PEG-PDMS-PEG Triblock copolymers

Sudipta Gupta[1*], Rasangi M. Perera[1*], Christopher J. Van Leeuwen[1], Tianyu Li[2], Laura Stingaciu[3], Markus Bleuel[4,5], Kunlun Hong[2,6] and Gerald J. Schneider[1,7*]

[1]Department of Chemistry, Louisiana State University, Baton Rouge, LA 70803, USA

[2]Department of Materials Science and Engineering, University of Tennessee, Knoxville, TN 37996, USA

[3]Neutron Sciences Directorate, Oak Ridge National Laboratory, POB 2008, 1 Bethel Valley Road, Oak Ridge, TN 37831, USA

[4]NIST Center for Neutron Research, National Institute of Standards and Technology, Gaithersburg, MD 20899-8562, USA

[5]Department of Materials Science and Engineering, University of Maryland, College Park, MD 20742-2115, USA

[6]Center for Nanophase Materials Sciences, Oak Ridge National Laboratory, Oak Ridge, TN 37831, USA

[7]Department of Physics & Astronomy, Louisiana State University, Baton Rouge, LA 70803, USA

## Abstract

Micelles and vesicles are promising candidates in targeted drug/gene delivery, bioreactors, and templates for nanoparticle synthesis. We investigated the morphology and dynamics of PEG-PDMS-PEG triblock copolymer nano-scale assemblies regarding the membrane dynamics because the molecular dynamics of the membrane govern mechanical properties like the stability of a membrane. We studied the structure by cryogenic transmission electron microscopy, small-angle neutron scattering, and the dynamics by dynamic light scattering and neutron spin echo spectroscopy. We changed the length of the hydrophilic block to obtain micellar and vesicular systems. The vesicle has a membrane rigidity, $\kappa_\eta = 16 \pm 2 \; k_B T$, the same order of magnitude as the corresponding liposome value but one order of magnitude higher than polymeric interfaces in microemulsions. Hence, the height-height fluctuations of polymers in a polymersome seem much less than those measured for surfactants at an oil-water interface. Therefore, the polymersome is substantially more stable. The value is very close to liposomes, indicating a similar stability.

# 1 Introduction

Amphiphilic triblock copolymers can self-assemble into different morphologies in aqueous solutions, including spherical micelles, rods, vesicles, disk-like, and other complex polymer aggregates such as giant compound micelles with an inverted core-shell structure surrounded by a hydrophilic surface. [1-8] Vesicles can further arrange from a simple spherical structure to elongated tubular or onion-like structures. [8, 9] Some of these nano-scale scale self-assembled structures have emerged as promising candidates as targeted drug/gene delivery agents, bioreactors, and templates for nanoparticle synthesis and to build cell-like structures and bioreactors. [4, 10-18] Therefore, it is clear that the scope, application, and usage of self-assembled structures of polymers are vast fields, and studying the fundamental nature of the structure and dynamics of these assemblies is very important.

Studies on the self-assembled structures of these nano-scale assemblies are found excessively in the literature. Polymer chain length, hydrophilic to hydrophobic ratio between different blocks, polymer architecture, and solvent nature profoundly impact the structures' size and shape. [19-22] Apart from those, external stimuli like salt concentration, temperature, and pH can influence the shape, size, and aggregation number transformations. [23-26] The impact of these parameters has been demonstrated using a vast range of techniques, mainly but not limited to dynamic light scattering (DLS), microscopy, small-angle neutron scattering (SANS), and small-angle X-ray scattering (SAXS). [13, 27-30] Often, these studies of morphology report shape deformation under external perturbations such as osmotic stress, temperature, and pH changes. [24-26, 31-33]

The microscopic dynamics of these objects govern the macroscopic material response. For example, the use of micelles and polymersomes as drug/gene carriers may be related to the stability of these objects, which can be expressed in terms of the nanoscopic fluctuations. [34-37] For biological membranes, changes in fluidity and composition of microdomains have been proven to regulate many critical physiological signaling pathways. [38] In other words, membranes exchange matter, energy, and information (signaling). Fluctuations on different length scales may lead to the formation of static or dynamic patches, so-called rafts, resulting in a heterogeneous state of matter in the membrane. [39, 40] These phenomena can be applied to polymer vesicles and micelles in drug/gene delivery since their membranes are the first parts that would be in contact with cellular/organelle membranes. Also, membrane thickness and rigidity affect the release rate of entrapped particles. [41] Considering all these facts, it can be stressed that the study of molecular motions in the polymer self-assemblies is essential to give an insight into the changes in the membrane dynamics.

Thus far, studies of the polymer dynamics of membranes are rare. However, the experience from vesicular liposomes (vesicles resulting from phospholipid self-assemblies) shows the importance of bending and stretching moduli, lysis tension, and pore edge tension. Connecting the moduli and dynamics of membranes is highly relevant for drug delivery systems, where the membrane resistance to deformation may determine the behavior of a membrane. Hence, bending rigidity, area compressibility, or stretching elasticity modulus may be critical parameters to characterize the response of membranes, even with the complexity of the surrounding biological environment. [42]

The key to understanding the dynamics of objects is morphology. To obtain objects of defined structures, we varied the amphiphilicity of the polymer. The self-assembled structure of a polymer can be explained by its hydrophilic-to-hydrophobic ratio. [8, 43, 44] As a rule of thumb, if the hydrophilic mass fraction of a copolymer lies between 25-45%, it usually forms vesicles. Suppose the hydrophilic mass fraction of a copolymer lies between 45-55%. In that case, it favors the formation of cylindrical micelles, and at even higher hydrophilic mass fractions, the formation of spherical micelles is favored. [44, 45] However, these ratios are not definitive features, and many exceptions can be found in the literature. [8]

Polymersomes with thicker and more rigid membranes than the lipid vesicle (liposome) membranes are expected to show higher bending rigidities, lysis tensions, and stretching elasticity values. [42] It is essential to directly measure the membrane dynamics to verify existing models, extend our knowledge beyond the capabilities of current theory or simulations, and develop next-generation soft materials.

This study used two PEG-PDMS-PEG triblock copolymers with different hydrophilic weight fractions, which self-assembled into spherical micelles and vesicles. Both PEG and PDMS blocks of the triblock polymer find widespread applications. For example, PEG is one of the unique polymers specifically used in medical and biological applications. [46, 47] Most of its biological applications include PEG-protein conjugates and PEG-modified small molecules for pharmaceutical applications, PEG tethers for biomolecule synthesis, PEG hydrogels for cell encapsulation, drug delivery, and wound covering, and PEG-containing liposomes and micelles for drug delivery. [48, 49] Its non-toxicity, non-immunogenicity, biocompatibility with blood and tissue, and solubility in a wide range of solvents offer various possibilities of utilization and explain the number of studies dealing with this polymer. [50, 51] It is also one of the few FDA-approved polymers (Food and Drug Administration, USA). [52, 53] Conversely, PDMS is widely used as an optically transparent, flexible, inert, nontoxic, non-flammable, biocompatible material. The low interacting methyl groups of the PDMS backbone cause bioinertness, which is why it has been routinely used as a biomedical implant material and for fundamental cellular studies. [54-57]

The bending rigidity of a membrane can be obtained by several techniques, including fluctuation spectroscopy, which is based on direct video microscopy observation of vesicles, methods based on mechanical deformation, molecular dynamics simulations, and Neutron Spin Echo (NSE) spectroscopy. [35, 58-62] NSE techniques have been critical for studying liposomal or modified liposomal membranes. [36, 62-64] The liposome results suggest that the NSE technique is beneficial for studying the dynamics of polymersomes.

This paper used DLS, SANS, and cryo-transmission electron microscopy (cryo-TEM) to characterize the static structure. NSE spectroscopy and DLS were used to study nanoscopic and mesoscopic dynamics to understand how the nanoscopic structure affects the nanoscopic molecular dynamics. NSE spectroscopy has been quite successful in exploring molecular motions in vesicles. [65-67]

## 2 Theory

### 2.1 Morphology

The 1D SANS scattering pattern for a unilamellar PEG-PDMS-PEG polymersome can be described by [23]

$$I(Q, R, t, \Delta\rho) = \frac{d\Sigma}{d\Omega}(Q) = \frac{\phi[A(Q)]^2}{V(r_3) - V(R_c)} \qquad (1)$$

Here $\phi$ is the block copolymer volume fraction. The scattering contribution from three different shells consisting of PEG, PDMS, and PEG are given by

$$A^2(Q) = A_1^2 + A_2^2 + A_3^2 + A_{12} + A_{23} + A_{13} \qquad (2)$$

with $A_{12} = 2A_1A_2$, $A_{12} = 2A_2A_3$, and $A_{13} = 2A_1A_3$ are the cross terms for PEG-PDMS, PDMS-PEG, and the two outer PEG-PEG layers, respectively. Here $r_3 = R_p = R_c + 2t_{PEG} + t_{PDMS}$ is the outer radius of the vesicle, with the radius of the core, $R_c$, the thickness of the PEG shells, $t_{PEG}$, and the thickness of the PDMS shell, $t_{PDMS}$. This model has been adapted for modified core-shell model used for vesicles. [68-70] A modified core-shell model used for micelles has also been adapted based on our previous studies. [71, 72] The 1-D SANS scattering pattern is given by:

$$I(Q) = \frac{d\Sigma}{d\Omega}(Q) = \frac{\phi}{V_m}[I_{core}(Q) + I^b_{corona}(Q) + I_{inter}(Q) + I_{blob}(Q)] \quad (3)$$

Here $V_m$ and $\phi$ are the micelles' volume and volume fraction, respectively. The terms $I_{core}(Q)$ and $I^b_{corona}(Q)$ correspond to the scattering contribution from the micellar core and corona, respectively. The third contribution, $I_{inter}(Q)$, comes from the interference terms between the core and the corona and the blob scattering, $I_{blob}(Q)$, from the swollen corona is implemented following Svaneborg and Pedersen. [73]

### 2.2 Dynamics

A simple approach to analyze the intermediate scattering function, $S(Q,t)$, as determined by NSE experiments, uses the Zilman-Granek (ZG) model that introduces the bending rigidity to describe the membrane dynamics [74]

$$\frac{S(Q,t)}{S(Q)} = \exp\left[-(\Gamma_Q t)^{2/3}\right]. \quad (4)$$

The only free parameter is the $Q$-dependent decay rate, $\Gamma_Q$, from which we derive the intrinsic bending modulus, $\kappa_\eta$, by [66, 75, 76]

$$\frac{\Gamma_q}{Q^3} = 0.0069\gamma \frac{k_B T}{\eta} \sqrt{\frac{k_B T}{\kappa_\eta}} \quad (5)$$

Here $\eta$ is the viscosity, $k_B$ the Boltzmann constant, $T$ the temperature, and $\gamma$ is a weak, monotonously increasing function of $\kappa_\eta/k_B T$. [74] In case of lipid bilayers usually $\kappa_\eta/k_B T \gg 1$, leading to $\gamma = 1$. [65, 66, 74, 76, 77] Equation 5 can be derived from a modified ZG theory that includes intermonolayer friction. A detailed discussion is omitted here but can be found in the literature. [66, 75, 76]

In addition, the contribution of the translational center of mass diffusion, $D_t$, of the liposomes needs to be included in the analysis of the dynamic structure factor, and we rewrite equation 1 [66, 78]

$$\frac{S(Q,t)}{S(Q)} = \exp(-D_t Q^2 t) \exp\left[-\left(\Gamma_q t\right)^{2/3}\right] \tag{6}$$

assuming ZG motion and center of mass diffusion are statistically independent. The separate measurement of $D_t$ by DLS avoids additional free parameters. Alternative concepts may use a stretched exponent Kohlrausch-Williams-Watts (KWW) function coupled to the diffusion, $D_t$ given by

$$\frac{S(Q,t)}{S(Q)} = \exp(-D_t Q^2 t) \exp\left[-\left(\frac{t}{\tau}\right)^{\beta}\right] \tag{7}$$

The KWW function is very popular in explaining the polymer dynamics in melts, solutions, and glassy materials. [79-82] This empirical approach has the advantage that it does not assume any particular concept and can provide information about the nature of the sub-diffusive behavior of the polymer chain exhibiting Zimm - or Rouse-type relaxation.

## 3 Experimental

**Sample Preparation.** The two amphophilic triblock copolymers were synthesized via hydrosilylation between hydride-terminated PDMS blocks and allyloxy (poly(ethylene oxide)) methyl ether blocks, as described in our previous work. [23, 83] The $PEG_{14}$-$PDMS_{15}$-$PEG_{14}$ and $PEG_{28}$-$PDMS_{15}$-$PEG_{28}$ polymers were dissolved in $D_2O$ and subjected to sonication for 30 seconds using a probe sonicator. To obtain monodisperse unilamellar vesicles, the resulted solution for the $PEG_{14}$-$PDMS_{15}$-$PEG_{14}$ polymer was extruded through a 0.1 µm pore size polycarbonate membrane using a double-syringe mini-extruder (Avanti Polar Lipids, Alabaster, AL, Canada).

**Dynamic Light Scattering (DLS)** measurements were performed using a Malvern Zetasizer Nano ZS equipped with a He-Ne laser of wavelength λ = 633 nm at 30 mW laser power, at a scattering angle θ = 173°.

**Cryogenic-transmission electron microscopy (cryo-TEM)** images were recorded in the bright field setting on a Tecnai G2 F30 operated at 150 kV and a JEOL 1230 operated at 40 – 120 kV.

**Small-angle neutron scattering (SANS)** experiments were conducted at the NG 7 SANS instrument of the NIST Center for Neutron Research (NCNR) at the National Institute of Standards and Technology (NIST). [84] The sample-to-detector distances, $d$, were fixed to 1, 4, and 13 m at neutron wavelength, $\lambda$ = 6 Å. Another configuration with lenses at $d$ = 15.3 m and $\lambda$ = 8 Å was used to access low $Q$-values. [85] This combination covers a $Q$ - range from ≈ 0.001 to ≈ 0.6 Å$^{-1}$, where $Q = 4\pi \sin(\theta/2)/\lambda$, with the scattering angle, $\theta$. A wavelength resolution of $\Delta\lambda/\lambda$ = 14% was used. All data reduction into intensity, $I(Q)$, vs. momentum transfer, $Q = |\vec{Q}|$ was carried out following the standard procedures implemented in the NCNR macros for the Igor software package. [86] The intensity values were scaled into absolute units (cm$^{-1}$) using the direct beam. The solvents and the empty cell were measured separately as backgrounds.

**Neutron spin echo spectroscopy (NSE)** was conducted at SNS-NSE BL15 at the Spallation Neutron Source of the Oak Ridge National Laboratory, Oak Ridge, TN. [87] We used Hellma quartz cells with a 2 mm sample thickness. Measurements were conducted at an incident wavelength of 8 Å. The BL15 Dr. Spine software package was used for data reduction.

## 4  Results and Discussion

Hereafter, we concentrate on two triblock copolymers based on poly(dimethyl siloxane) (PDMS) and poly(ethylene glycol) blocks, PEG$_{14}$-PDMS$_{15}$-PEG$_{14}$ and PEG$_{28}$-PDMS$_{15}$-PEG$_{28}$. The index indicates the number of repeating units in each block.

As illustrated in Figure 1, we have calculated the electric field autocorrelation function, $g_e^1(Q,t)$, which for identical interacting spheres gives the normalized dynamic structure factor, $S(Q,t)/S(Q)$. The data has been modeled using a simple diffusion model, $\exp(-D_t Q^2 t)$, to obtain the translational diffusion coefficient, $D_t$. The hydrodynamic radius, $R_h$, of each self-assembled structure was calculated using the Stokes-Einstein equation, $R_h = k_B T/(6\pi \eta_0 D_t)$, with the Boltzmann constant, $k_B$, the temperature, $T$, and the viscosity of the solvent (D$_2$O), $\eta_0$ = 1.0945 cP. [88] D$_2$O has a lower background than H$_2$O in SANS and NSE experiments. Hence, we used D$_2$O also for DLS.

We obtain $R_h$ = 59.5 ± 0.5 nm for the PEG$_{14}$-PDMS$_{15}$-PEG$_{14}$ polymersome, with $D_t =$ (3.37 ± 0.03) × 10$^{-12}$ m$^2$s$^{-1}$. For PEG$_{28}$-PDMS$_{15}$-PEG$_{28}$ micelles, we obtain $R_h$ = 8.02 ± 0.07 nm, with $D_t =$ (24.9 ± 0.2) × 10$^{-12}$ m$^2$s$^{-1}$.

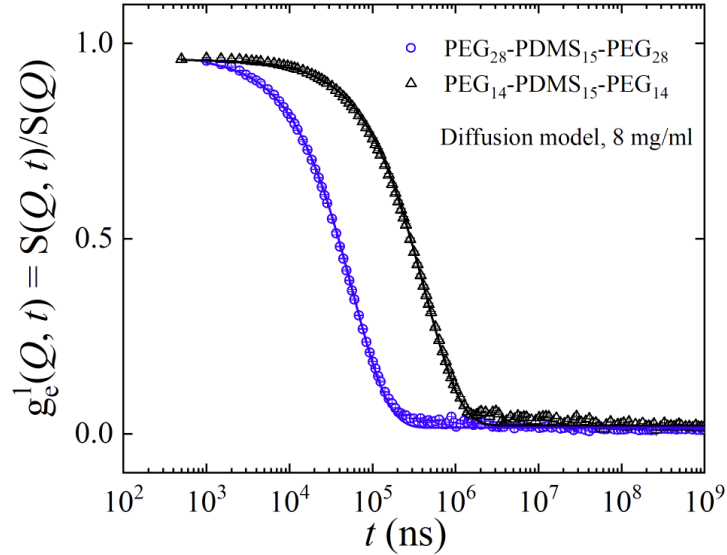

Figure 1. The electric field autocorrelation function, $g_e^1(Q,t)$, from DLS as a function of lag time for PEG$_{28}$-PDMS$_{15}$-PEG$_{28}$, and PEG$_{14}$-PDMS$_{15}$-PEG$_{14}$ tri-block copolymer samples at 8 mg/ml concentrations. The solid lines represent the fits using a simple diffusion model, as explained in the text.

Figure 2a illustrates the SANS scattering intensity normalized by their volume fraction, ϕ, for the PEG$_{28}$-PDMS$_{15}$-PEG$_{28}$ and PEG$_{14}$-PDMS$_{15}$-PEG$_{14}$ polymers. The results for PEG$_{14}$-PDMS$_{15}$-PEG$_{14}$ are modeled using a vesicle form factor, equation 1. [69, 70] Our SANS analysis yields the total number of polymers per polymersome, $N_{agg}$ = 65954 ± 128, with the PEG and PDMS thickness of 0.6 ± 0.2 nm and 5 ± 0.8 nm, respectively. These values correspond to a PEG$_{14}$-PDMS$_{15}$-PEG$_{14}$ membrane thickness of 5.6 ± 1.0 nm. The corresponding outer perimeter radius $R_{SANS} = R_p$ = 59.2 ± 2.0 nm with a particle polydispersity of 30 ± 2 %, obtained from a log-normal size distribution.

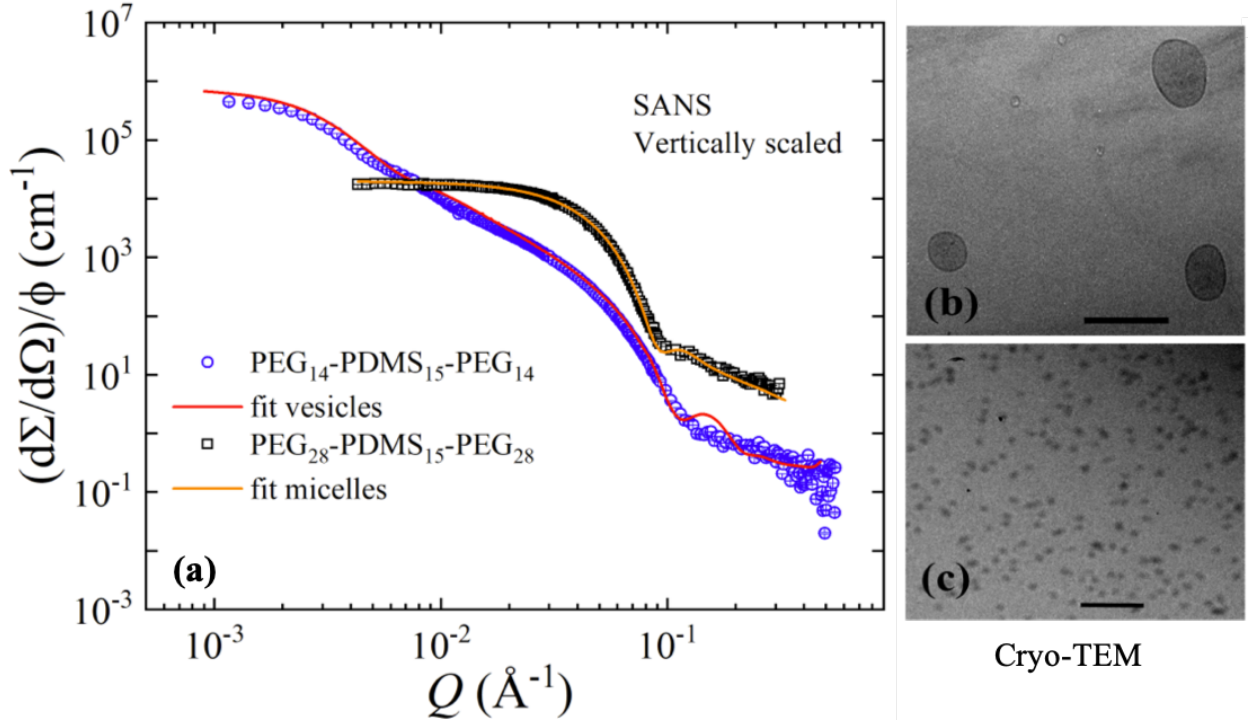

*Figure 2. (a) SANS scattering intensity normalized by their volume fraction, ϕ, for PEG$_{14}$-PDMS$_{15}$-PEG$_{14}$, and PEG$_{28}$-PDMS$_{15}$-PEG$_{28}$ tri-block copolymer samples at 8 mg/ml concentrations dispersed in D$_2$O. The solid lines represent the fits using a vesicle and a micelle model. Cryo-TEM images of self-assembled structures of (b) PEG$_{14}$-PDMS$_{15}$-PEG$_{14}$ and (c) PEG$_{28}$-PDMS$_{15}$-PEG$_{28}$ with scale bars representing 200 and 50 nm, respectively.*

Instead, the data for the PEG$_{28}$-PDMS$_{15}$-PEG$_{28}$ polymer have been modeled using the micelle form factor, equation 3. [71, 72] We obtain a micellar radius, $R_{SANS} = R_m = 5.1 \pm 0.7$ nm. The micellar core radius is given by $R_c = 3.4 \pm 0.1$ nm, where the standard deviation of a Gaussian distribution, $\sigma_G = 0.19 \pm 0.01$, has been used to define the corona-solvent interface of the micelle. The corresponding aggregation number, i.e., the number of polymers per micelles, $N_{agg} = 100 \pm 4$.

Figure 2b illustrates the cryo-TEM images of the PEG$_{14}$-PDMS$_{15}$-PEG$_{14}$ polymer that forms unilamellar polydisperse vesicles with the largest diameter of around 200 nm and the smallest diameter of around 88 nm. The average membrane thickness of the vesicles was $6 \pm 1$ nm. As illustrated in Figure 2c, PEG$_{28}$-PDMS$_{15}$-PEG$_{28}$ polymer self-assembled into micelles with an average diameter of 3.5 nm.

The macromolecular dynamics ultimately determine the structural stability. Hence, in our next step, we explore the dynamics of the $PEG_{14}$-$PDMS_{15}$-$PEG_{14}$ and $PEG_{28}$-$PDMS_{15}$-$PEG_{28}$ triblock copolymers.

Figure 3a illustrates the normalized dynamic structure factor, $S(Q,t)/S(Q)$, covering a $Q$-range from 0.035 to 0.094 Å$^{-1}$, measured by NSE for the vesicular sample ($PEG_{14}$-$PDMS_{15}$-$PEG_{14}$). The data can be modeled using the ZG (equation 5) and the KWW models (equation 7), as illustrated by the solid and dashed lines. The translational diffusion, $D_t$, from DLS, was kept as a fixed parameter. The corresponding stretched exponent obtained from the KWW fits using equation 7 was $\beta = 0.59 \pm 0.02$. This $\beta$ value is very close to 0.66 (= 2/3), as predicted by Zilman and Granek. The calculated curves are very similar. Overall, the ZG equation describes the experimental results well and yields a membrane rigidity of $\kappa_\eta = 16 \pm 2\ k_BT$, as obtained from Figure 4.

We also point to two famous models, Rouse dynamics ($\beta \approx 0.5$) of polymers in melts and Zimm-like motion ($\beta \approx 0.66$) for polymers in solution. [79, 89] Recent studies on poly(styrene)-poly(butadiene) (PS-PB) block copolymer micelles suggested melt-like behavior of PB hydrophobic core in n-hexadecane and DMF. [90] This, along with our analysis, opens up the possibilities for a Zimm-Rouse-like motion exhibited by the triblock copolymer in a selective solvent.

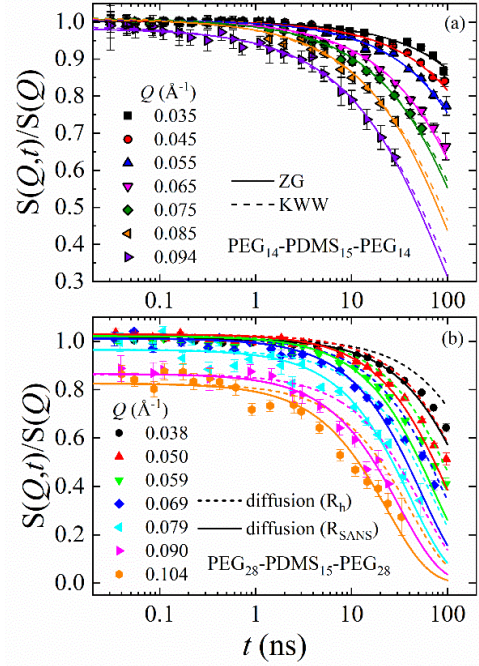

*Figure 3. Dynamic structure factor, S(Q,t)/S(Q), as a function of Fourier time, t, for different Q's for PEG$_{14}$-PDMS$_{15}$-PEG$_{14}$ and PEG$_{28}$-PDMS$_{15}$-PEG$_{28}$ tri-block copolymer samples at 8 mg/ml concentrations dispersed in D$_2$O. The vesicle data in (a) is modeled using* the ZG (*equation 5*) *and* KWW (equation 7) models, *presented by the solid and the dashed lines, respectively. The micelle data in (b) is compared to the translational diffusion,* $\exp(-D_t Q^2 t)$, *of a rigid sphere. For illustration, the translational diffusion coefficient,* $D_t$, *has been calculated from the hydrodynamic radius,* $R_h$, *(dashed lines), and the micellar perimeter radius,* $R_m$, *(solid lines).*

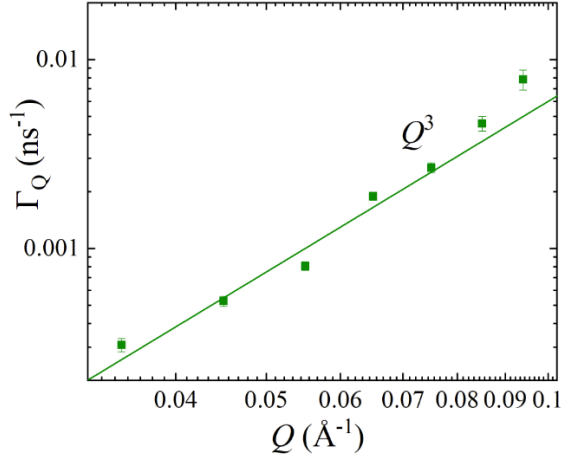

*Figure 4. Zilman-Granek (ZG) parameter, $\Gamma_Q$, vs. momentum transfer, Q. The full line represents calculations with Equation 5.*

Figure 3b illustrates the normalized dynamic structure factor, $S(Q,t)/S(Q)$, measured by NSE for the micellar sample (PEG$_{28}$-PDMS$_{15}$-PEG$_{28}$), covering a Q-range from 0.038 to 0.104 Å$^{-1}$. The comparison of data modeling following the translational diffusion, described by the diffusion coefficient $D_t = k_B T/(6\pi\eta_0 R_{h/m})$ corresponding to hydrodynamic radius, $R_h$, from DLS and the micellar radius, $R_m$, from SANS has been performed. The good agreement of the model with the experimental results indicates that the micelles' observed dynamics structure factor S(Q,t)/S(Q) of the micellar sample is governed by translational diffusion.

## 5 Conclusion

We investigated the molecular structure and dynamics of the two amphiphilic triblock copolymers, PEG$_{14}$-PDMS$_{15}$-PEG$_{14}$ and PEG$_{28}$-PDMS$_{15}$-PEG$_{28}$. Small-angle neutron scattering and cryo-transmission electron microscopy confirmed the formation of vesicles and micelles. Neutron spin echo spectroscopy experiments show substantially different results, with a vesicle that can be well described by Zilman-Granek and Zimm dynamics and a micelle well represented by the translational center of mass diffusion. Hence, there is a substantial difference between the two different assembled structures. The membrane rigidity of $\kappa_\eta = 16 \pm 2\ k_B T$ (at ambient temperature) is the same magnitude as the value measured for many liposomes. For example, depending on temperature, pH, and buffer conditions, DOPC is 20 k$_B$T. [34, 35] Instead, it is much bigger than values observed for microemulsion with polymeric surfactants, typically in the order of 1 k$_B$T. [91, 92] Although different materials have similar membrane-thickness since the lipid and polymer-based vesicles, the observed bending rigidity is closer to the liposome values. It has been observed that the membrane rigidity approximately varies quadratically with an increase in thickness. [93] Hence, it is likely that polymer-based systems allow a wide variety of bending elasticities, from the low values found at interfaces at microemulsions to values that match those of vesicles based on lipids. In understanding the dynamics, these unique polymersomes are established as a model macromolecule that can bridge the gap between self-assembled micro-emulsions and liposomes. This is a first step forward in the fundamental understanding of the structure-morphology-dynamics of such a hybrid-macromolecular system, and further investigations are essential to determine the concentration dependence and influence of the chemical composition.

# 6 Author Information

## 6.1 Corresponding Authors

*Email: Sudipta Gupta g.sudipta26@gmail.com

*Email: Rasangi Perera mpere4@lsuhsc.edu

*Email: Gerald J. Schneider gjschneider@lsu.edu

## 6.2 ORCID

Sudipta Gupta: 0000-0001-6642-3776

Rasangi Perera: 0000-0002-1394-4006

Laura Stingaciu: 0000-0003-2696-5233

Markus Bleuel: 0000-0003-3923-9505

Kunlun Hong: 0000-0002-2852-5111

Gerald J. Schneider: 0000-0002-5577-9328

# 7 Notes

The authors declare no competing financial interest.

# 8 Acknowledgment

The neutron scattering work is supported by the U.S. Department of Energy (DOE) under EPSCoR Grant No. DE-SC0012432 with additional support from the Louisiana Board of Regents. The Center for High-Resolution Neutron Scattering provided access to the neutron spin echo spectrometer and small-angle scattering instruments, a partnership between the National Institute of Standards and Technology and the National Science Foundation under Agreement No. DMR-1508249. Research conducted at the Spallation Neutron Source (SNS) at Oak Ridge National Laboratory (ORNL) was sponsored by the Scientific User Facilities Division, Office of Basic Energy Sciences, U.S. DOE. The polymer synthesis and characterization of this research was conducted at the Center for Nanophase Materials Sciences (CNMS), which is a DOE Office of Science User Facility. The Center for High-Resolution Neutron Scattering provided access to the NG7 SANS instrument, a partnership between the National Institute of Standards and Technology and the National Science Foundation under Agreement No. DMR-201079. We want to thank Jeff Krzywon for NG7 beamline support. We want to thank Rafael Cueto (LSU) for supporting DLS experiments and Jibao He (Tulane University) for his support in Cryo-TEM imaging.